\documentclass[preprint,proceedings]{rmaa}

\def\hMpc{\ifmmode{h^{-1}{\rm Mpc}}\else{$h^{-1}{\rm Mpc}$}\fi}
\def\hMsun{\ifmmode{h^{-1}M_\odot}\else{$h^{-1}M_\odot$}\fi}

\SetYear{2002}
\SetConfTitle{Galaxy Evolution: Theory and Observations}

\title{Low mass dark matter halos in voids}

\author{
  S. Gottl\"ober,\altaffilmark{1}
  E. {\L}okas\altaffilmark{2}
  and A. Klypin,\altaffilmark{3}}

\altaffiltext{1}{Astrophysikalisches Institut Potsdam.}
\altaffiltext{2}{Copernicus Center, Warsaw.}
\altaffiltext{3}{New Mexico State University.}

\shortauthor{Gottl\"ober, {\L}okas \& Klypin}
\shorttitle{Halos in voids}

\fulladdresses{
\item Stefan Gottl\"ober: Astrophysikalisches Institut Potsdam,
An der Sternwarte 16,
14482 Potsdam, Germany.
\item Ewa {\L}okas: Nicolaus Copernicus Astronomical Center, Bartycka 18,
00-716 Warsaw, Poland.
\item Anatoly Klypin: Astronomy Department,
New Mexico State University, Dept.\ 4500,
Las Cruces, NM 88003, USA.
}

\listofauthors{S. Gottl\"ober, E. {\L}okas \& A. Klypin}
\indexauthor{Gottl\"ober, S.}
\indexauthor{{\L}okas, E.}
\indexauthor{Klypin, A.}

\abstract{Using numerical simulations and the Sheth-Tormen
approximation we study the mass function of dark matter halos in
voids. We find that the void mass function is significantly lower and
its shape is different than that of the field halos. We predict that
in the standard LCDM model a void with radius $10\hMpc$ should have 50
halos with circular velocity $v_c>50$ km/s and 600 halos with $v_c>20$
km/s.}

\addkeyword{Large--scale structure of Universe}
\addkeyword{Cosmology: theory}

\begin{document}
\maketitle

According to the hierarchical structure formation paradigm halos form
from small initial density fluctuations by accretion and merging. It
is expected that inside the dark matter (DM) halos gas cools and
galaxies are formed. Thus, we expect that the distribution of DM halos
and their structure are closely related to the observed galaxy
distribution. It is also known (e.g., Gottl\"ober et al. 2001) that
the merging history of the DM halos depends on environment. Here we
study DM halos in the low density environment of voids.

Cosmological simulations predict many more small DM halos than the
observed number of satellites around the Milky Way and Andromeda
galaxies.  Do we have the same problem for dwarf galaxies in voids?
One naively expects a large number because the Press-Schechter mass
function steeply rises with declining mass. In contrast, it seems that
observations are failing to find a substantial number of dwarf galaxies
inside voids (e.g. Popescu et al. 1997).  However, the situation is
complicated because it is very difficult to detect dwarf galaxies,
many of which are expected to have low surface brightness.

We use N-body simulations to make accurate predictions for the
expected number of dwarf halos in voids. We have performed a series of
high resolution simulations using the ART code (Kravtsov et al. 1997)
and the friends-of-friends (FOF) algorithm to identify halos.  We
investigate a spatially flat cold dark matter model with a
cosmological constant ($\Lambda$CDM with $\Omega_{\rm M}=0.3$,
$\Omega_{\Lambda}=0.7$, $\sigma_8=0.9$, and $h=0.7$).  We use a cube
of 80\hMpc\ side length, which is sufficiently large to study the
formation of large voids.  Because we are interested in the formation
of small structure elements inside the voids, we need high mass
resolution. Therefore, we perform the simulation in two steps. First,
we run a low mass resolution simulation ($128^3$ particles with mass
$m_{\rm part} = 2.0 \times 10^{10}\hMsun$) and identify 8387
galactic-size halos with masses $>2.0 \times 10^{11} \hMsun$. This
corresponds to a mean distance of about $4 \hMpc$ between halos.  We
identify voids in the distribution of the halos. The largest void has
a diameter of $24 \hMpc$. Second, we re-simulate the voids with a
formal mass resolution of $1024^3$ particles, i.e. $m_{\rm part} = 4.0
\times 10^7 \hMsun$. Thus, we resolve in the voids objects with masses
larger than $10^9 \hMsun$. By construction, the voids do not have
halos with masses larger than $M_{\rm b} = 2.0 \times 10^{11} \hMsun$.

We resimulate five voids with radii $R_{\rm void} = 11.6, 10.8, 9.4,
9.1, 9.1 \ \hMpc$. The underdensity at the centers of these voids is
0.1 -- 0.2 of the mean matter density $\Omega_{\rm M}$.  In Fig. 1 we
compare the mass function of DM halos in voids with the mass function
in the whole box. The thin solid line is the analytical prediction
based on the Sheth and Tormen (1999) correction to the 
Press-Schechter mass function
\vspace{-0.04in}
\begin{eqnarray}
    n_{\rm ST}(M) &=& - \left( \frac{2}{\pi} \right)^{1/2}
    A \left[ 1 + \left( \frac{a \delta_c^2}{\sigma^2} \right)^{-p}
    \right] a^{1/2} \frac{\varrho_{\rm b}}{M}       \nonumber  \\
    & \times & \frac{\delta_{\rm c}}{\sigma^2} \frac{{\rm d} \sigma}{{\rm d} M}
    \exp \left( - \frac{a \delta_{\rm c}^2}{2 \sigma^2} \right),  \label{v11}
\end{eqnarray}
where $A=0.322$, $p=0.3$, $a=0.707$, $\varrho_{\rm b}$ is the
background density, $\sigma$ is the rms density fluctuation on scale
$R$ corresponding to mass $M$ ($M= 4 \pi \varrho_{\rm b} R^3/3$) and
$\delta_{\rm c}=1.676$ for our model.  As expected, within the whole
mass range covered by the numerical mass function it agrees well with
the analytical prediction. Small deviations at $M > 10^{14} \hMsun $
are due to low number of massive objects.

\begin{figure}[!t]
  \includegraphics[width=\columnwidth]{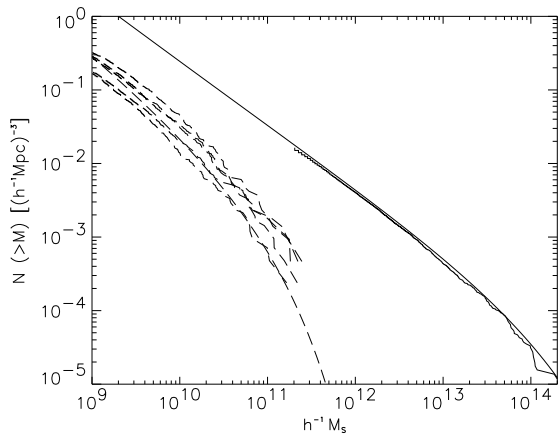}
  \caption{Mass function of DM halos.  The thick full line shows the
cumulative mass function $N(>M)$ in the whole $80 \hMpc$ simulation.
The thin full curve is the Sheth-Tormen approximation.  The mass
functions in voids are shown by thick dashed curves. The thin dashed
curve is our Sheth-Tormen-type prediction of the void mass function.
The mass functions of the voids are different
because of the difference in the void underdensity.  Note that the
normalization {\it and the shape} of the void functions are different
from that of the field population.}
  \label{fig:massf}
\end{figure}

The simulated voids have somewhat different mean (under)densities.
This results in different mass
functions. The number-density of halos in voids is about an order of
magnitude smaller than in the whole box as expected due to
lower mean density in voids. However, the shape of the mass
function is  also different.
An analytic approximation to mass function
in voids, shown as a thin dashed line in Fig. 1, was obtained by applying
the formalism of Sheth and Tormen (2002) for the constrained mass functions
\adjustfinalcols
\begin{eqnarray}
    n_{\rm c, ST}(M) &=& - \left( \frac{2}{\pi} \right)^{1/2}
    \frac{\varrho_{\rm void}}{M}
    \frac{| T(\sigma^2 | \sigma_0^2)|}{(\sigma^2 -\sigma_0^2)^{3/2}}
    \frac{\sigma {\rm d} \sigma}{{\rm d} M}  \nonumber  \\
    & \times & \exp \left[ - \frac{[B(\sigma^2) - B(\sigma_0^2)]^2}
    {2 (\sigma^2-\sigma_0^2)} \label{v14}
    \right]
\end{eqnarray}
where
\begin{displaymath}
    T(\sigma^2 | \sigma_0^2) = \sum_{n=0}^5 \frac{(\sigma_0^2 - \sigma^2)^n}{n!}
    \frac{\partial^n [B(\sigma^2) - B(\sigma_0^2)]}{\partial (\sigma^2)^n}
\end{displaymath}
and
\begin{eqnarray}
    B(\sigma^2) &=&
    a^{1/2} \delta_c[1+ \beta (a \delta_c^2/\sigma^2)^{-\alpha}],   \nonumber   \\
    B(\sigma_0^2) &=&
    a^{1/2} \delta_0 [1+ \beta (a \delta_0^2/\sigma_0^2)^{-\alpha}].   \nonumber
\end{eqnarray}
The parameter $\delta_0$ is the linear underdensity of the void
corresponding to the actual nonlinear underdensity $\delta$ and is
calculated from the spherical top-hat model, while $\sigma_0$ is the
rms fluctuation at scale $R_0 = (1+\delta)^{1/3} R_{\rm void}$.  The
parameters $\alpha=0.615$, $\beta=0.485$ come from the ellipsoidal
dynamics and we have chosen $a=0.5$ which fits the simulated mass
functions better than the value of $a=0.707$ advertised by Sheth and
Tormen. The result shown in Fig. 1 was obtained for $\Omega_{\rm
void}=0.04$ and $R_{\rm void}=10 \hMpc$.  In an upcoming paper
(Gottl\"ober et al. 2002, in preparation) we discuss the spatial
distribution of halos found in numerical simulations of voids and
their mass function, and we present a straightforward analytical
method to predict the mass function.

In a typical void of radius $10 \hMpc$ we found about 600 halos with
circular velocities larger than 20 km/s and about 50 halos with
circular velocities larger than 50 km/s. Testing these
predictions could be one of the interesting observational tasks of
the ongoing SLOAN Digital Sky Survey.

\end{document}